\documentclass[aps,showpacs,floatfix,superscriptaddress,prd,11pt]{revtex4-1}
\usepackage{graphicx}
\usepackage{amsmath}
\usepackage{amssymb}
\usepackage{bbm}
\usepackage{hyperref}
\usepackage{color}
\usepackage[utf8]{inputenc}
\usepackage{CJKutf8}

\newcommand{\eref}[1]{(\ref{#1})}
\newcommand{\nn}{\nonumber}
\newcommand{\sign}{\rm sign}
\newcommand{\be}{\begin{eqnarray}}
\newcommand{\ee}{\end{eqnarray}}

\newcommand{\bmat}{\left ( \begin{array}{cc} }
	\newcommand{\emat}{\end{array} \right ) }

\newcommand{\beq}{\begin{equation}}
\newcommand{\beqs}{\begin{equation*}}
\newcommand{\eeq}{\end{equation}}
\newcommand{\eeqs}{\end{equation*}}

\allowdisplaybreaks

\begin{document}

\title{Universality and Thouless energy in the supersymmetric Sachdev-Ye-Kitaev Model}
\author{Antonio M. Garc\'\i a-Garc\'\i a}
\email{amgg@sjtu.edu.cn}
\affiliation{Shanghai Center for Complex Physics, School of Physics and Astronomy,
	Shanghai Jiao Tong University, Shanghai 200240, China}
\author{Yiyang Jia\begin{CJK*}{UTF8}{gbsn}
(贾抑扬)
\end{CJK*}}
\email{yiyang.jia@stonybrook.edu}
\affiliation{Department of Physics and Astronomy, Stony Brook University, Stony Brook, New York 11794, USA}

\author{Jacobus J. M. Verbaarschot}
\email{jacobus.verbaarschot@stonybrook.edu}
\affiliation{Department of Physics and Astronomy, Stony Brook University, Stony Brook, New York 11794, USA}

\begin{abstract}
  We investigate the supersymmetric Sachdev-Ye-Kitaev (SYK) model, $N$ Majorana fermions with infinite range interactions in $0+1$ dimensions. We have found that,
  close to the ground state $E \approx 0$, discrete symmetries alter qualitatively the spectral properties with respect to the
non-supersymmetric SYK model.
The average spectral density at finite $N$, which we compute analytically and numerically, grows exponentially with $N$ for $E \approx 0$. However the chiral condensate, which is normalized with respect the total number of eigenvalues, vanishes in the thermodynamic limit. 
Slightly above $E \approx 0$, the spectral density grows exponential with the energy. 
Deep in the quantum regime, corresponding to the first $O(N)$ eigenvalues,
the average spectral density is universal and well described by random matrix ensembles with chiral and superconducting discrete symmetries. The dynamics for $E \approx 0$ is investigated by level fluctuations.
Also in this case we find excellent agreement with the prediction of chiral and superconducting
random matrix ensembles for eigenvalues separations smaller than the Thouless
energy, which seems to scale linearly with $N$. Deviations beyond the Thouless energy, which describes how ergodicity is approached, are universality characterized by a quadratic growth of the number variance. 
In the time domain, we have found analytically that the spectral form factor $g(t)$, obtained from the connected two-level correlation function of the unfolded spectrum, decays as $1/t^2$ for times shorter but comparable to the Thouless time with $g(0)$ related to the
coefficient of the quadratic growth of the number variance.
Our results provide further support that quantum black holes are ergodic and therefore can be classified by random matrix theory. 
 
\end{abstract}

\maketitle
\section{Introduction}
Random matrix theory \cite{wigner1951,dyson1962a,dyson1962b,dyson1962c,dyson1962d,dyson1972} is a powerful tool to explain universal features of complex quantum systems. In general, it is applicable for long times scales where the system has equilibrated and therefore its motion depends only on global symmetries, such as time reversal invariance and charge conjugation.  
Quantitative agreement with random matrix theory indeed have been reported in a variety of problems: from disordered and quantum chaotic systems  in the limit of negligible localization effects \cite{bohigas1984} to the spectrum of highly excited nuclei and Lattice Quantum Chromodynamics (QCD) \cite{shuryak1993,verbaarschot1993,verbaarschot1994}. In the latter, it was found that both the density and level statistics of the low-lying eigenvalues of the QCD Dirac operator follow the predictions of random matrix theory where the choice of the ensemble, that labels the universality class, depends on the representation of the gauge group and the number and flavors. Despite its simplicity, random matrices can capture important dynamical features of realistic strongly correlated systems, for instance chiral symmetry breaking, a salient feature of non-perturbative QCD, related to the infrared limit of the spectral density \cite{banks1980} of the QCD Dirac operator.

Despite its success, the random matrix approach has obvious limitations as for sufficiently short times where the quantum dynamics is not universal. In certain cases, such as non-interacting disordered metals and quantum chaotic systems, where there is a good understanding of these corrections, it is possible to estimate the time or energy scale, usually called Thouless energy or time, where non-universal corrections are expected and to even get quantitative information on the dynamics in this intermediate scale by studying the corrections to the random matrix results \cite{altshuler1986}.
For strongly interacting systems, the situation is less clear though several approaches have been developed to tackle this problem in the context of condensed  matter \cite{altshuler1997}, QCD \cite{osborn1998} and nuclear physics \cite{french1970,french1971,bohigas1971,bohigas1971a,kota2014}.

In the latter, fermionic models with infinity range two-body interactions were introduced  as a natural generalization of random matrix models with the expectation of a larger range of applicability while keeping some analytical control.
More recently, research on this problem has received an important new impetus after the realization \cite{kitaev2015} that $N$ 
fermions with infinite range two-body interactions, now called Sachdev-Ye-Kitaev (SYK) model \cite{kitaev2015,sachdev2010,sachdev1993}, could be a toy model for holography, namely,
the infrared limit be could dual to a certain quantum black hole in an anti-deSitter space in two bulk dimensions ($AdS_2$). 

Basic features of the SYK model, that support the existence of a
gravity dual \cite{maldacena1998},
includes, maximal chaos
\cite{maldacena2015,kitaev2015,maldacena2016,jensen2016,polchinski2016,maldacena:2016upp}
in the strong coupling limit,
finite zero temperature entropy \cite{kitaev2015,parcollet1998,georges2001,sachdev2010}
linear specific heat in the low temperature limit \cite{Jevicki:2016bwu,Jevicki:2016ito},
exponential growth of the low energy excitations \cite{garcia2017,cotler2016,Bagrets:2016cdf,Bagrets:2017pwq} and short range spectral correlations given by random matrix theory \cite{garcia2016,cotler2016,krishnan2017a}. More recently, different extensions of the SYK model have been studied including several fermionic flavors \cite{gross2017,rosa2017},
higher dimensions
\cite{davison2017,jian2017,Turiaci:2017zwd,Das:2017pif,Das:2017wae}, one plus two-body interactions \cite{garcia2017b,song2017,Chen:2017dav}, with supersymmetry
\cite{fu2017,Peng:2016mxj,Peng:2017spg,li2017,kanazawa2017,hunter2017,Narayan:2017hvh,murugan2017} and with no disorder \cite{Witten:2016iux,Klebanov:2016xxf,krishnan2017a} 

In these supersymmetric SYK models, global symmetries, that depend  on the number of Majoranas, agree \cite{li2017,kanazawa2017} with those of the chiral and superconducting \cite{altland1997} ensemble of random matrices \cite{verbaarschot1994}. Since global symmetries of the
non-supersymmetric model match those of the standard Wigner-Dyson ensembles, there is a one-to-one correspondence between $N$ dependent SYK global symmetries and most of the known \cite{altland1997} universality classes
in random matrix theory. 

This raises several questions: do SYK models with different
global symmetries still keep most of the features expected
in a theory with gravity-dual such as quantum chaos,
exponential increase of low energy excitations or a finite entropy at zero temperature? What is the precise window of universality in which random matrix theory is applicable and how are corrections beyond this universality regime?

In this paper we aim to answer these questions. We study the spectral density,
thermodynamic properties and level statistics in a supersymmetric SYK model by analytical and exact diagonalization techniques. Our main results are that
this chiral SYK model still has all the expected features of a gravity-dual: finite entropy at zero temperature, exponential increases of low energy excitations and excellent agreement of level statistics with random matrix theory predictions for sufficiently long time or small energy separations. We also show that not only level statistics but also the spectral density in the infrared limit is universal and given by random matrix theory. We study quantitatively deviations from these universal results, observed for energies (times) larger (smaller) than the so called Thouless energy (time) \cite{altshuler1986}. The number variance grows quadratically for eigenvalue separations larger than the Thouless energy and the spectral form factor decays as a power-law for times shorter than the Thouless time. 

The paper is organized as follows. 
In section II, we start with the definition of the model and then discuss analytical 
results for the spectral density and spectral resolvent of the supercharge and the thermodynamic properties of the Hamiltonian. In section III we compare numerical results for the microscopic spectral density with random matrix predictions. Level statistics and its  comparison
with random matrix theory, including deviations for (small) large (time) energies, are
discussed in section IV. Concluding remarks are made in section V. In Appendix
A we work out the large $N$ limit of the Q-Hermite result for the spectral density while a simple analytical form for  the resolvent of the SYK model is obtained in Appendix B.
 
 \section{Model, average spectral density and thermodynamic properties}

 \subsection{The SYK Model}
 
 We investigate a SYK model, namely $N$ Majorana fermions with infinite range $q$-body interactions, for $q=3$.
 The Hamiltonian is given by the square of the supercharge $Q$
 \cite{fu2017}
\be
H =Q^2,
\ee
where the supercharge is defined by,
  \begin{equation}\label{hami}
 Q \, = \, i\sum_{i,j,k=1}^N J_{ijk} \, \gamma_i \, \gamma_j \, \gamma_k \, ,
 \end{equation}
 with $\gamma_i$ Majorana fermions defined by the following algebraic relation
 \begin{eqnarray}
 \{ \gamma_i, \gamma_j \} = \delta_{ij},\label{clif}
 \end{eqnarray}
 also verified by Euclidean Dirac $\gamma$ matrices. 
  The coupling  $J_{ijk}$ is chosen to be a Gaussian random variable 
 with probability distribution,
 \begin{equation}
 P(J_{ijk}) \, = \, \sqrt{\frac{N^2}{36\pi J^2}} \exp\left( - \, \frac{N^2J_{ijk}^2}{36J^2} \right) \, ,
 \end{equation}
 where $J$ sets the scale of the distribution. Unless specified otherwise $J \equiv 1$.
 
 This Hamiltonian is supersymmetric with global symmetries that, depending on $N$, correspond to one of the chiral and superconducting
 random matrix ensembles \cite{fu2017,kanazawa2017}. 
 In this paper we will study the thermodynamical properties of the Hamiltonian,
 but we will analyze the spectral properties of the supercharge for $E \sim 0$ corresponding to the
 ground state and low energy excitations of the Hamiltonian.

  \begin{figure}[t!]
 	\includegraphics[width=10.cm]{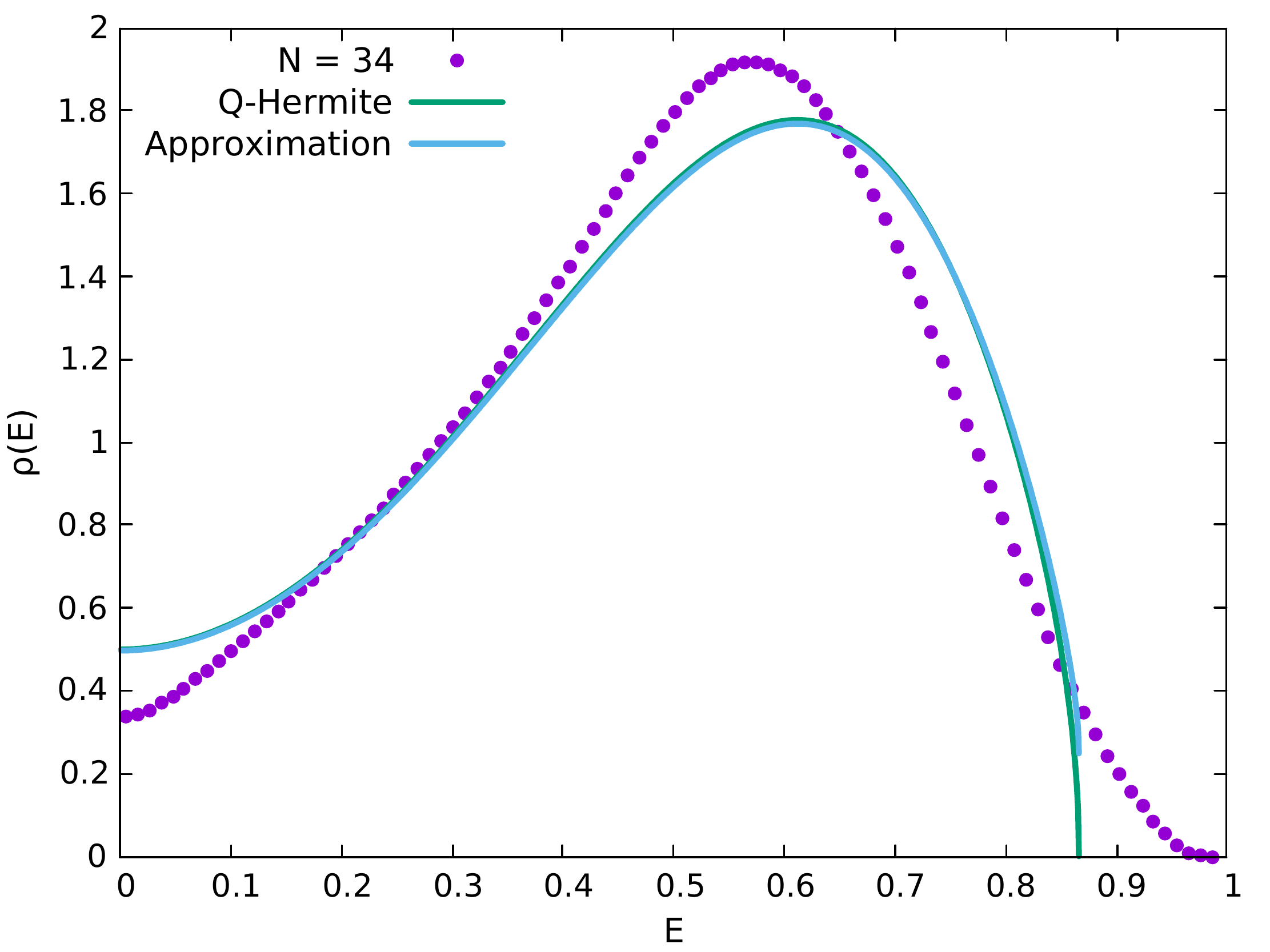} 
 	\caption{Average spectral density of the supercharge $Q$ (with Hamiltonian $H = Q^2$) of the SYK model Eq.~(\ref{hami}) for $q=3$ and $N=34$.
          We find excellent agreement between analytical Q-Hermite result Eq.~(\ref{dqher}) and the approximation Eq.~(\ref{rhoasym}). However, agreement with the exact diagonalization results (solid dots), employing $5\times10^6$ eigenvalues, is only qualitative.
 	}
 \label{dgcoma}
 \end{figure}
 
  \subsection{The Spectral Density of the Supercharge}
  
  The average spectral density for even $q$, computed in Ref. \cite{garcia2017} by an explicit evaluation of the moments, turned out to be given by the weight function of the  $Q$-Hermite polynomials with $Q = \eta$ where $\eta(N,q)$ is a suppression factor related to the  commutation of two  products of $q$ Majorana  operators. 
  A similar calculation can be carried out for odd $q$. 
The average density is still given by the  same analytical expression but with a negative suppression parameter (for even $q$, $\eta$ can also become negative
for small values of $N$),  
\be
\eta_ = {N \choose q}^{-1}  \sum_{r=0}^q  (-1)^{r+q}{q \choose r} {N-q \choose q-r}.
\label{suppress}
\ee
 In terms of $|\eta|$ the average density can be rewritten as 
 \be
 \rho(E) &=& c_N \sqrt{1-\frac{E^2}{E_0^2 }}\prod_{k=1}^\infty
 \left[1-4\frac{E^2}{E_0^2}\frac 1{2+|\eta|^{2k}+|\eta|^{-2k}} \right ]
 \left[1-4\frac{E^2}{E_0^2}\frac 1{2-|\eta|^{2k-1}-|\eta|^{-2k+1}} \right ].
 \hspace*{0.5cm}
 \label{dqher}
 \ee
Note that this expression is valid for both even and odd $q$.
After a Poisson resummation, an explicit evaluation of the resulting integrals is possible (see Appendix \ref{app:A}).
In the large $N$ limit, it simplifies to,
 \be 
 \rho_{\rm asym} (E) = c_N  \cosh \left( \frac {\pi\arcsin(E/E_0)}{\log|\eta|}\right)\exp\left [ 2 
 \frac { \arcsin^2(E/E_0)}{\log|\eta|} \right ]
 \label{rhoasym}
 \ee
provided that  the absolute values of the energy is away from $E_0$  given by
\be
E_0 = \frac{4 \sigma^2}{1-\eta},
\ee
where
\be
\sigma^2 = {N \choose q}\sigma_0^2
\ee
with $\sigma_0$ the standard deviation of $J$.
 For even $q$, this asymptotic expression turned to be not only an excellent approximation of the Q-Hermite result Eq.~(\ref{dqher}) but also of the exact density obtained by numerical diagonalization of Eq.~(\ref{hami}). 
Results depicted in Fig.~\ref{dgcoma} confirm that Eq.~(\ref{rhoasym}) is still an excellent approximation of the $Q$-Hermite results. However, the agreement with the spectral density from exact diagonalization for $N = 34$ is only qualitative. Since correction terms to the moments in Q-Hermite
approximation  Eq.~(\ref{dqher}) are of order $q/N$ 
the excellent agreement for $q$ even was unexpected.

\subsection{The resolvent}

In this section, we study the resolvent of the supercharge of the SYK model
which is an alternative way to investigate its average spectral properties.
 It is defined by
\be
G(z) = \sum_{\lambda_k} \frac 1{z+\lambda_k},
\label{res}
\ee
where the $\lambda_k$ are the eigenvalues of $Q$. The spectral density is
given by the discontinuity of the resolvent across the real axis. Conversely,
the resolvent Eq.~\eref{res} follows by  integrating over the spectral density.
Because the spectrum of  $Q$ in  Eq.~(\ref{hami}) 
 is symmetric under $\lambda_k \to -\lambda_k$, the resolvent on the imaginary axis 
 \be\label{resd}
 iG(is) = \frac 1{\cal N}\sum_{\lambda_k>0} \frac {2s}{\lambda_k^2+s^2}
 \ee
 is purely imaginary (so that $iG(is)$ is real).
 According to the Banks-Casher relation \cite{banks1980}, the resolvent in the $s \to 0$ and thermodynamic limits is directly related to the spectral density at $E = 0$.
  In the context of QCD, a finite value in this limit signals
  the spontaneous breaking of chiral symmetry, one of its most relevant low energy features.
  In supersymmetric SYK model, 
  it is unclear whether this interpretation is exactly applicable.
  However, it is still of interest to study it as a finite spectral density at
  the origin is a strong indication of a highly entangled ground state.

  Moreover, we have derived a compact analytical expression based on the known \cite{ismail1987,erdos2014} spectral density for Q-Hermite polynomials. We leave the details to the appendix \ref{app:res} and state here the final result \cite{ismail1987}, 
 \be \label{resa}
 iG(is) &=& \frac {\sqrt{1-\eta}}\sigma \sum_{k=0}^\infty 
 \frac { \eta^{k(k+1)/2}}
 {\left (\frac{s\sqrt{1-\eta}}{2\sigma}
 	+\sqrt{ \frac {s^2(1-\eta)}{4\sigma^2}+1} \right )^{2k+1}}.
 \ee

\begin{figure}[t!]
 	\includegraphics[width=12.cm]{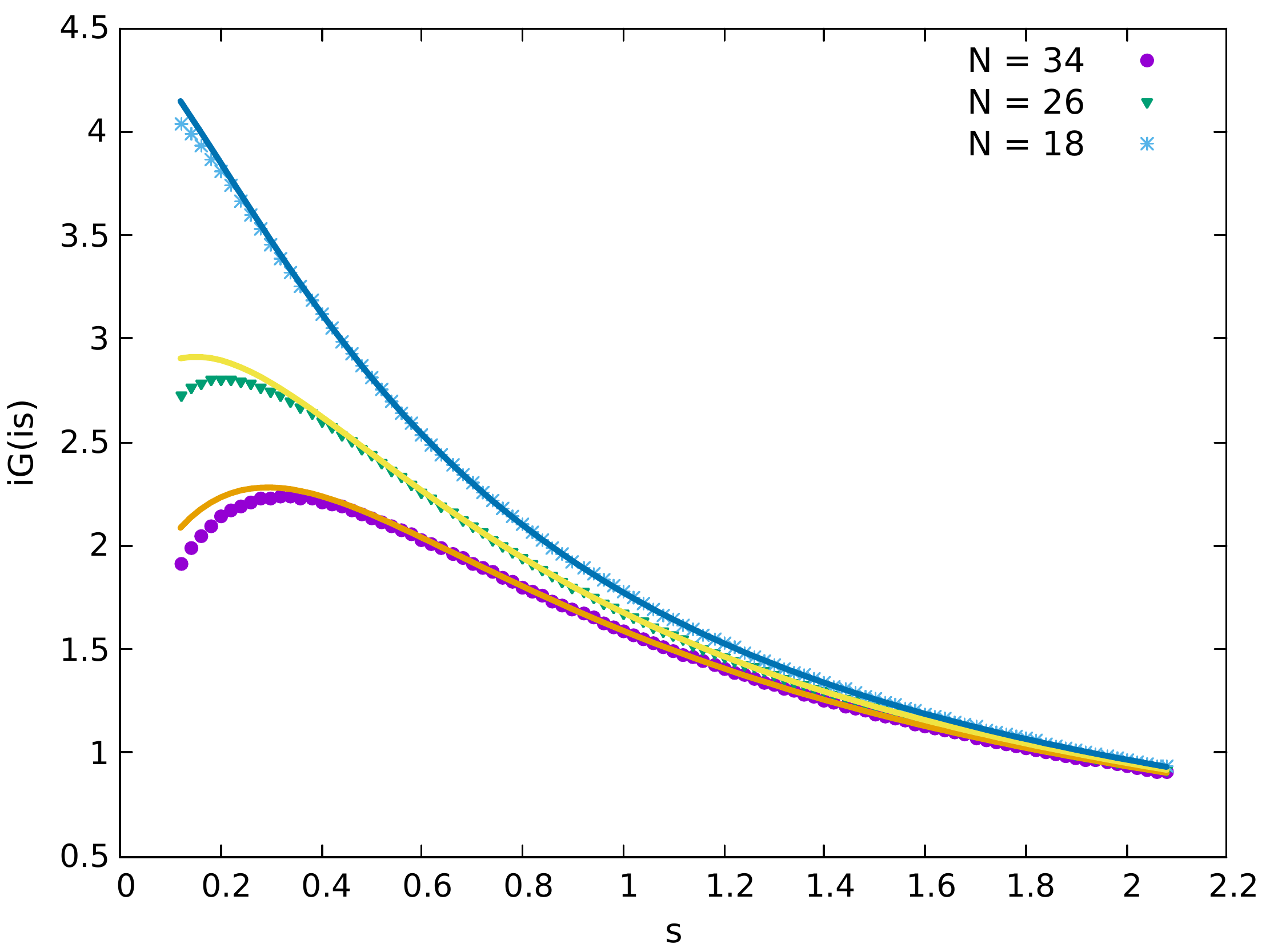}
 	\caption{Resolvent $iG(is)$ Eq.~(\ref{resd}) for the SYK model for $q=3$ as a function of the parameter $s$
          for  different value of $N$ compared with the analytical result Eq.~\eref{resa}. After a finite size scaling analysis, we find that the $N$-dependence of  $\log G(0) $ is well fitted by $\log G(0) \approx  -0.0695 N$ in agreement with the value of the q-dependent part of the
          zero temperature entropy \eref{entropy}.
We find excellent agreement with the analytical expression Eq.~(\ref{resa}) based on the Q-Hermite density except for very small $s$. 
 	}
 	\label{ccon}
 \end{figure}
 In Fig.~\ref{ccon} we depict results for the resolvent as a function of $s$ for different values of $N$. We find excellent agreement with the analytical prediction Eq.~(\ref{resa}) except for small $s$, dominated by high moments, where we expect the Q-Hermite result to be less accurate. 
 In order to explore numerically the limit $s \to 0$, $N \to \infty$ we carry a finite size scaling analysis in $N$ for a small $s \sim 0.01$. 
 We fit the $N$ dependence of the logarithm of the  resolvent by a function
 $a + b N$ where $a$ and $b$ are fitting parameters. The slope is equal to
 the nontrivial part of the entropy density. We find $b \approx -0.0695$ which agrees
 with the analytical result $-\pi^2/(16 q^2)$ to be discussed in the next subsection.

 \subsection{Thermodynamical properties}
In this subsection  we study the thermodynamical properties of the model in the large $N$
 and low temperature limit where the asymptotic average density
 Eq.~(\ref{rhoasym}) is expected to be a good approximation.
 We note that
 because the Hamiltonian $H = Q^2$, where $Q$ is the supercharge,
 the low temperature limit is controlled by states with positive
 energy close to $E = 0$. In contrast with even $q$ where for large
 $N$, $E_0 \propto N$, for $q$ odd when $\eta <0$, $E_0 \sim \sqrt N$.
 The region close to the origin for odd  $q$ case  is therefore qualitative different from the even $q$ case
 where the ground state energy scales with $N$ as should be the case for a fermionic system.
    
 The entropy at zero temperature $s_0$ follows from the 
 the prefactor $c_N$ in Eq.~\eref{dqher}.
 Although this constant is known analytically, see Appendix A,
 its large $N$ limit can simply be determined from the condition
 \be
 \int dE \rho_{\rm asym}(E) dE = 2^{N/2}.
 \ee
 Evaluating the integral by a saddle point approximation and using
 that in this limit $\log| \eta| \sim  -2 q^2/N$ we find
 \be
 c_N = e^{N/2\log 2- N\pi^2/16 q^2}.
 \ee
 Since the Q-Hermite approximation only has a nontrivial large $N$ limit
 when $q^2/N$ is kept fixed, it can only yield a $1/q^2$ correction to the
 zero temperature entropy which is given by
 \be
 s_0 = \log c_N = N\left (\frac 12 \log 2 - \frac{\pi^2}{16 q^2}\right ).
\label{entropy}
 \ee
 This result agrees up to order $1/q^2$ 
 with the exact result \cite{fu2017} of $s_0$ obtained by using path integral techniques. 
 We note that $s_0$ is larger than the $s_0$ in the  even $q$ case, where the nontrivial contribution to the entropy is $\pi^2/4 q^2$.
 This larger entropy, which suggests a highly entangled ground state, is likely a consequence of the special properties at $E \approx 0$ due to the chiral symmetry of the model. 
 
 As a further confirmation of this, we calculate the low temperature limit of the free energy from the asymptotic spectral density
 Eq.~\eref{rhoasym}. We show that it agrees with the large $q$ limit of the free energy for odd $q$ obtained in \cite{fu2017}.

 Since the Hamiltonian $H = Q^2 $ and $\rho_{\rm asym}(E)$ is the eigenvalue density
 of $Q$, the partition  function is given by
 \be
 Z(\beta) = \int dE \rho_{\rm asym}(E)e^{-\beta E^2}.
 \ee
 For large $N$, the integral can be evaluated by a saddle point approximation with
 saddle point equation given by
 \be
 -\frac \pi{E_0 \log|\eta|} \frac 1{\sqrt{1- (E/E_0)^2}} + \frac 4{E_0 \log|\eta|}
 \frac{\arcsin(E/E_0)}{\sqrt{1-(E/E_0)^2}}-2\beta E = 0.
\label{saddleE}
 \ee
 The free energy is therefore given by,
 \be
 -\beta F &=& \frac N2 \log 2 +\frac {\pi^2}{8\log|\eta|}
 -\frac{\pi\arcsin(E/E_0)}{\log|\eta|}+ 2\frac{\arcsin^2(E/E_0)}{\log|\eta|} -\beta E^2.
\ee
Substituting
 \be
 E = E_0 \sin( (\pi-v)/4)
 \ee
and using the saddle point equation 
 \be
 \frac {-v}{E_0^2 \log|\eta|
 	\cos(v/2)}- \beta = 0.
 \ee
 the free energy can be rewritten as
 \be
-\beta F &=& \frac N2 \log 2 + \frac{v^2}{8\log|\eta|} +\frac v{2\log |\eta|} \tan((\pi-v)/4).
 \ee
 If we identify $-E_0^2\log |\eta| = {\cal J}$ this is exactly the saddle point equation
 obtained in \cite{fu2017}. Taking
the large $N$ limit, it gives the free energy obtained in \cite{fu2017}
 \be
 -\beta F&=& 
 \frac N2 \log 2 - N\frac{v^2}{16 q^2} 
 - N\frac{\beta {\cal J}}{4q^2}+ N\frac{v}{2q^2}\tan v/2
 \ee
 with the constant energy $N{{\cal J}}/{4q^2}$
 subtracted.
 In the low-temperature limit, the solution of the saddle point equation is
 given by $v=\pi$ resulting in the free energy
 \be
 -\beta F&=& 
\frac N2 \log 2 - N\frac{\pi^2}{16 q^2} 
 - N\frac{\beta {\cal J}}{4q^2}+ O(1/\beta).
 \ee
 
 The conclusion of this analysis is
 that the Q-Hermite form of the spectral density which at large $N$ only
 has a nontrivial limit when $q^2/N$ is kept fixed reproduces
 the leading $1/q^2$ correction to the free energy. Contrary to the even $q$ case,
for odd $q$ the low-temperature limit is determined by energies close to zero. In the following section, we focus on this region only. 

\begin{figure}[t!]
 	\includegraphics[width=7.2cm]{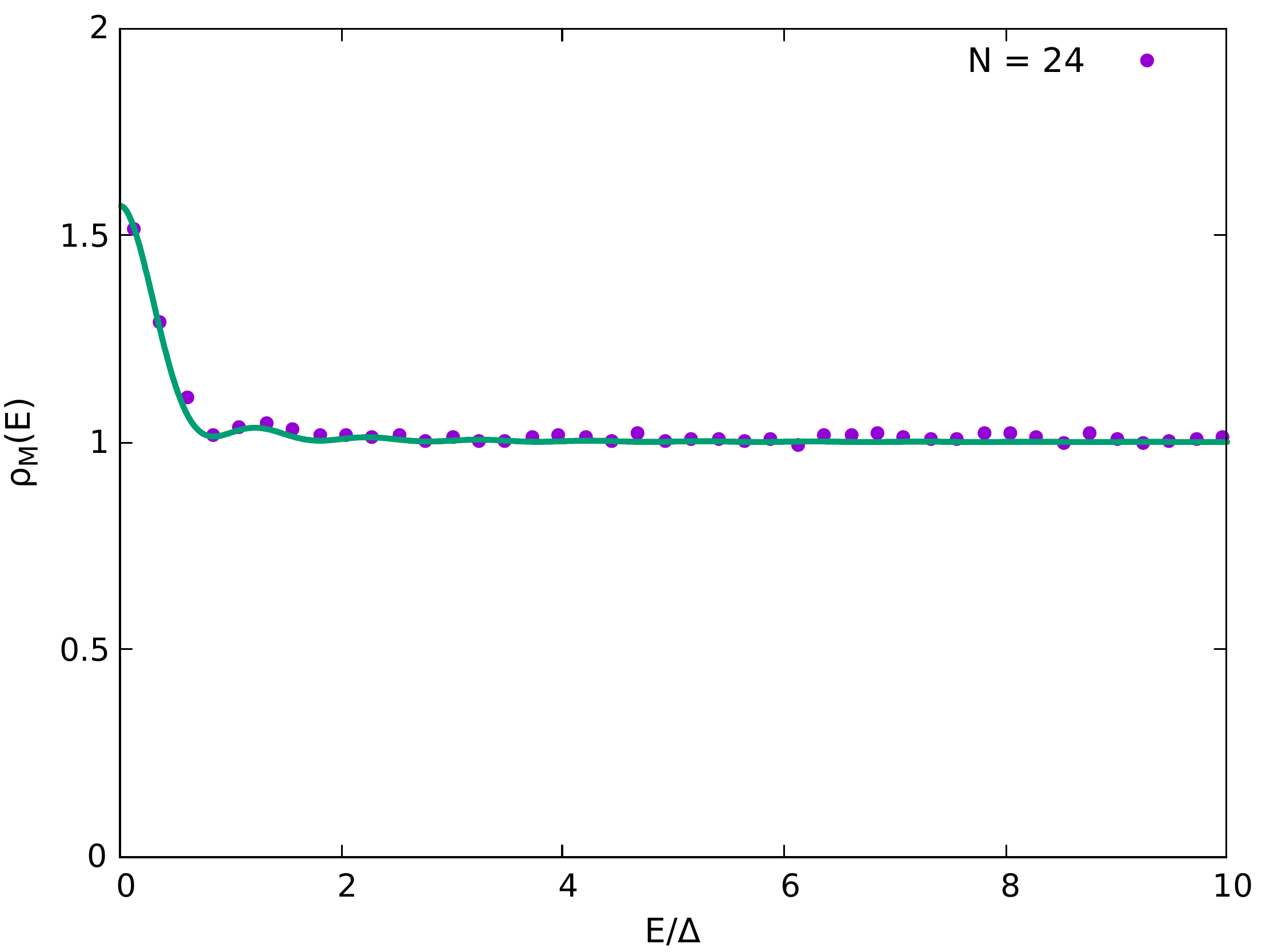}
 	\includegraphics[width=7.2cm]{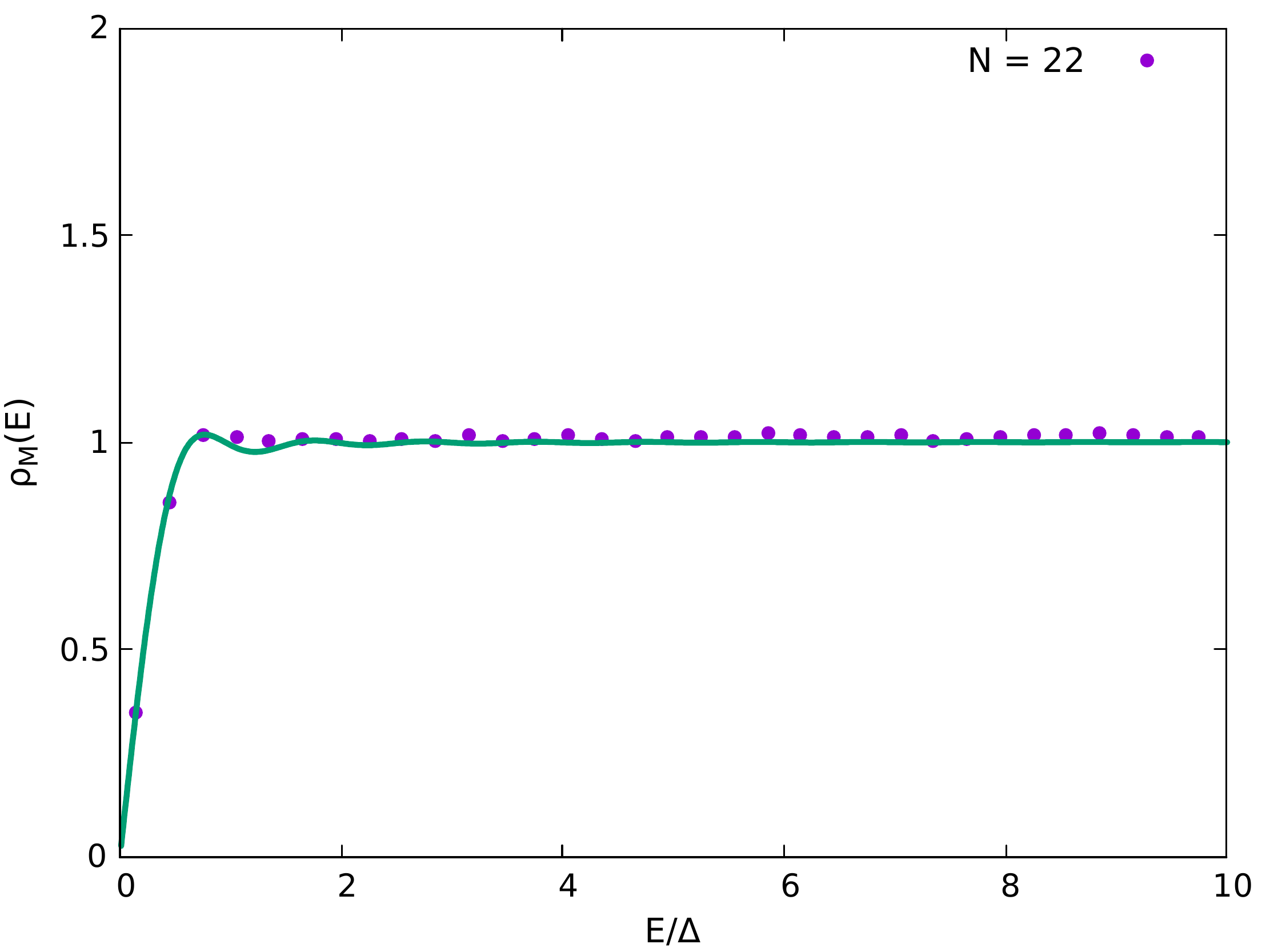}
 	\includegraphics[width=7.2cm]{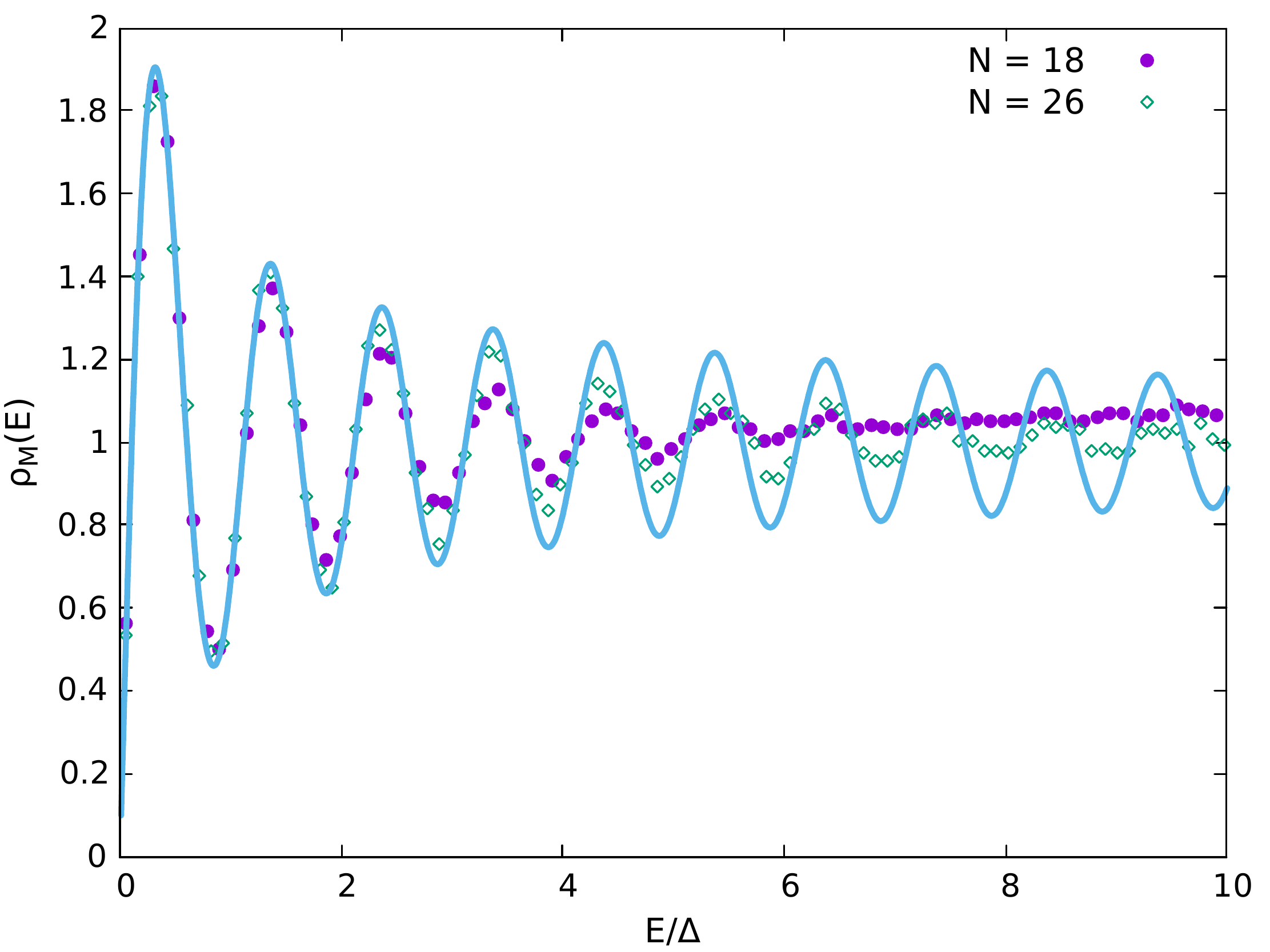} 
 	\includegraphics[width=7.2cm]{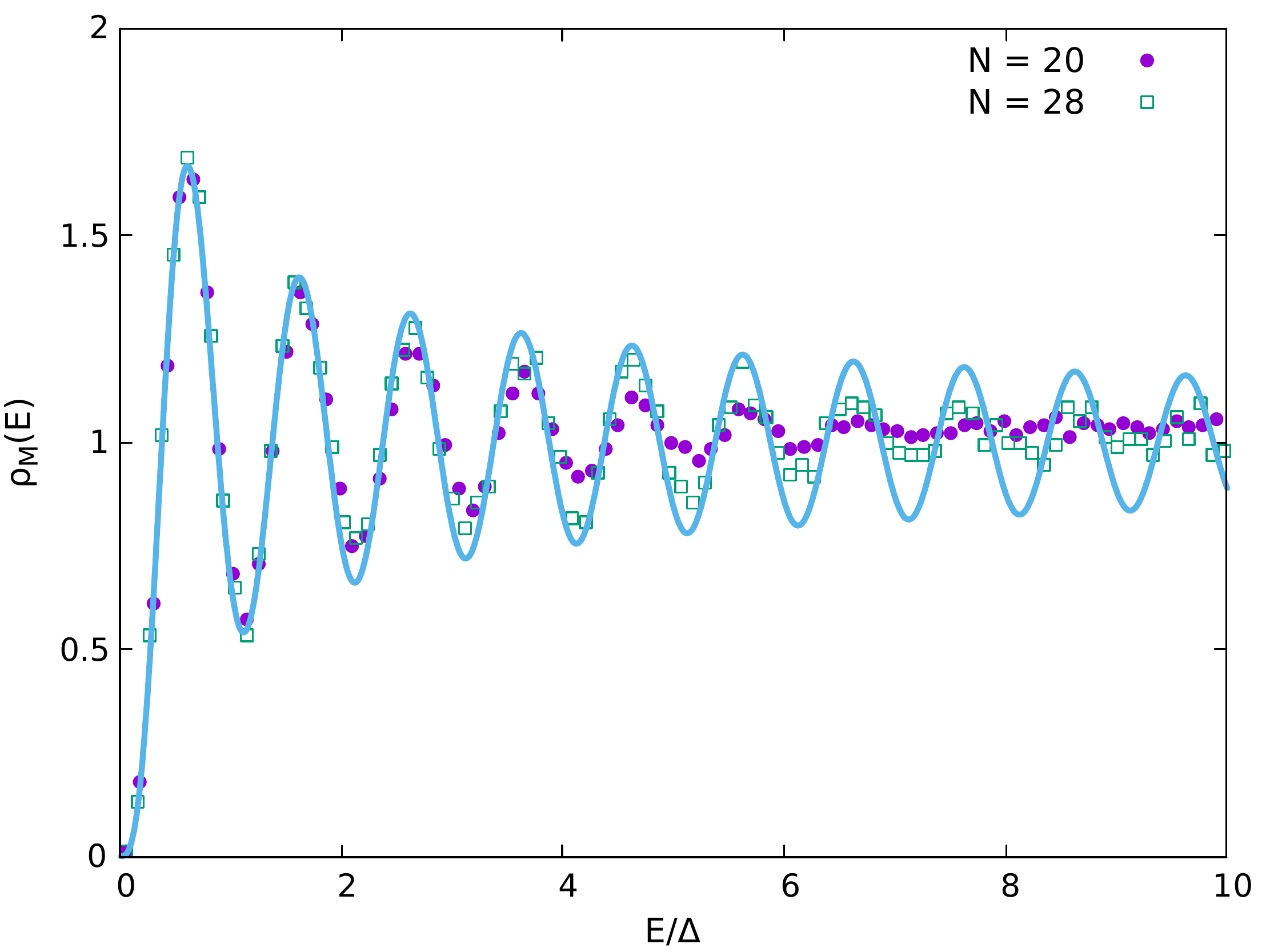} 
 	\caption{
          Microscopic spectral density $\rho_M$ near $E=0$ in units of the mean level spacing $\Delta$. We compare $\rho_M$ for the supercharge of
          the SYK model with $q=3$, see Eq.~(\ref{hami}), with the predictions
          of random matrix theory (solid curves).
          As was shown in \cite{li2017}, SYK models with different values of
          $N$ have different discrete global symmetries and therefore must be
          compared to random matrix ensembles belonging to the corresponding universality classes. For instance, $N = 18$ and $N=26$
          belongs to the $DIII$ ensemble, $N = 20$ and $N=28$ to the chiral symplectic ensemble (chGSE), $N = 24$ to the chiral orthogonal ensemble
          (chGOE), and $N = 22$ to the $CI$ ensemble. Interestingly, we observe an excellent
          agreement with the random matrix prediction for the first few
          low lying eigenvalues. The oscillations for the microscopic spectral density
          of the SYK model in the case of  $DIII$ and chGSE case (bottom row), are eventually washed out as the effect of
          discrete symmetries is strong only very close to $E =0$. Analytical expressions for all random matrix ensembles can be found in Refs.~\cite{verbaarschot1993,nagao1995,ivanov2002}. 
 	}
 	\label{denm}
 \end{figure}

 \section{Universal microscopic spectral density and resolvent}
 Depending on $N$, the supersymmetric SYK model \cite{fu2017,li2017} has
 additional chiral and discrete symmetries. The analysis of
 thermodynamical properties has revealed that the main differences with
 respect to the original SYK model is in the region close to the ground state
 $E \approx 0$. It is well known \cite{verbaarschot1993} that,
 precisely in this region, the microscopic spectral density $\rho_M(E)$
 defined by
 \be
 \rho_M(E) = \Delta \rho\left ( \frac E\Delta \right )
 \ee
 with $\Delta$ the level spacing near $E=0$,
 is universal and given by random matrix theory.

 Explicit
 analytical expressions are known for all universality classes
 \cite{nagao1995,verbaarschot1993,ivanov2002}. This is in stark contrast with
 Wigner-Dyson ensemble where only the spectral correlations, not the density,
 at the scale of the mean level spacing have universal features. The origin of
 this universality lies in the additional rigidity of the spectrum
 at $E \approx 0$ imposed by chiral symmetry which requires to the nonzero
 eigenvalues to occur in pairs $\pm \lambda_k$.
 In realistic systems, like
 QCD, this is only observed deep in the quantum regime corresponding
 to the first $O(F_\pi^2 \sqrt V)$ eigenvalues with $F_\pi$ the pion
 decay constant and $V$ the space-time volume. For practical  lattice QCD  simulations this amounts to only the first
 few eigenvalues above $E = 0$. Since the supersymmetric SYK model has
 the same global symmetries as those of the mentioned random matrix ensembles,
 it is natural to inquire whether the spectral density is also universal. 
 
 For that purpose, we compute the spectral density by exact diagonalization
 techniques. For any given $N$, we calculate
 at least $8\times 10^7$ eigenvalues.
Specifically, for $N=26,\;30,\;32,\;34$, the number of disorder realizations is $27340,\;4430,\;1300$ and $540$, respectively.
 Such a large number of eigenvalues
  is necessary to suppress statistical fluctuations. The SYK microscopic spectral density for the first few eigenvalues, depicted
  in Fig.~\ref{denm}, shows and excellent agreement with random matrix ensemble belonging to the
 same universality class. We note that this is a completely parameter-free comparison.
  
 For the chiral symplectic ensemble, we observe a suppression of the
 oscillations  for eigenvalue sufficiently far from the origin. This is
 consistent with results from other  systems such as QCD
 \cite{verbaarschot1993} where dynamical features not present in the
 random matrix ensemble tend to soften the effect of discrete symmetries.
 The scale at which the oscillations disappear, known as the Thouless
 energy, increases with $N$, and although an estimate of the precise $N$ dependence is hard,
 it seems that it scales linearly with $N$. 
 We note the distribution of the first eigenvalue was already worked out
 in Ref.~\cite{kanazawa2017} where  agreement
 with random matrix theory was also found.\\ 

     \begin{figure}[t!]
     	\includegraphics[width=7.2cm]{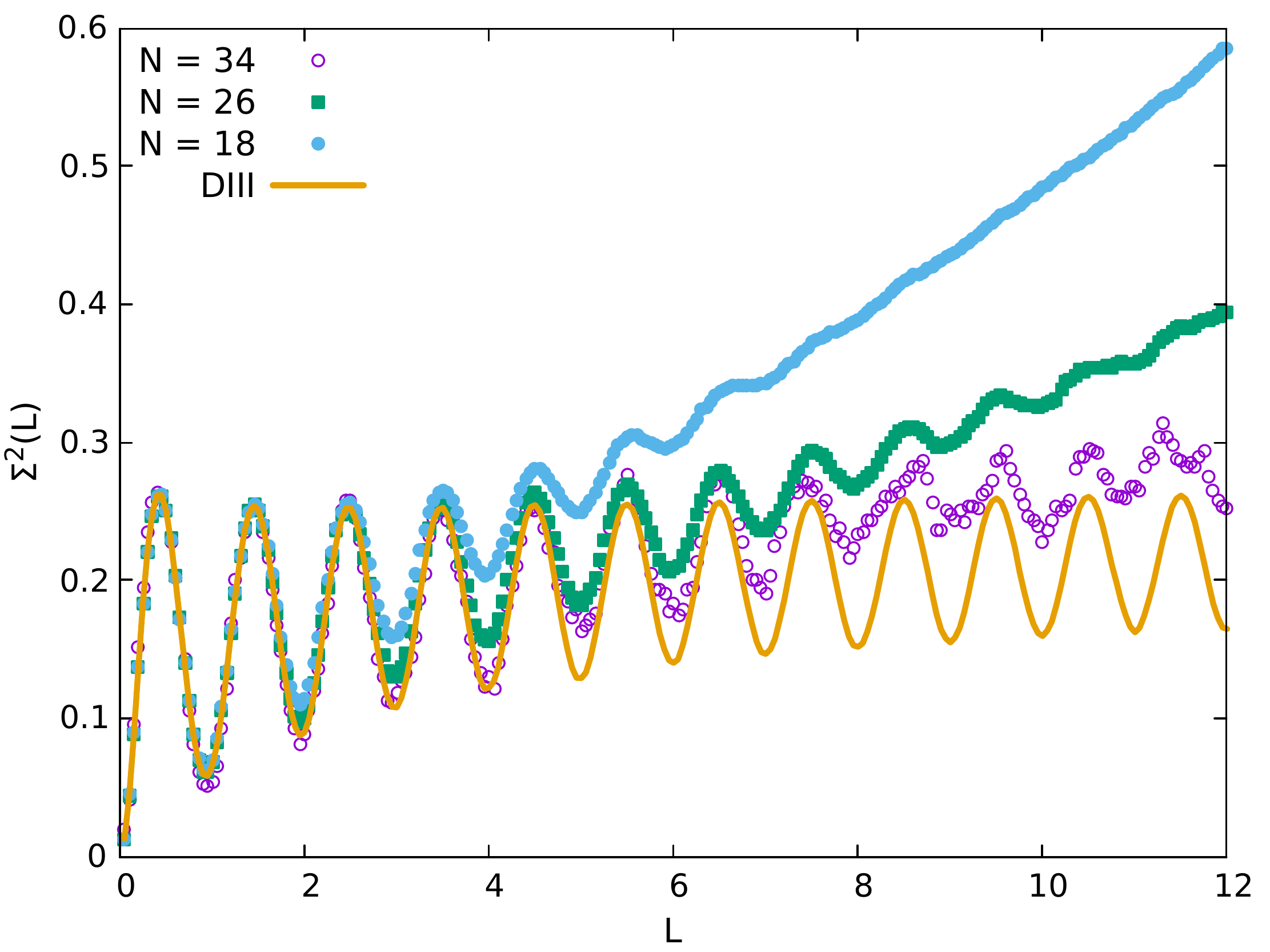}
     	\includegraphics[width=7.2cm]{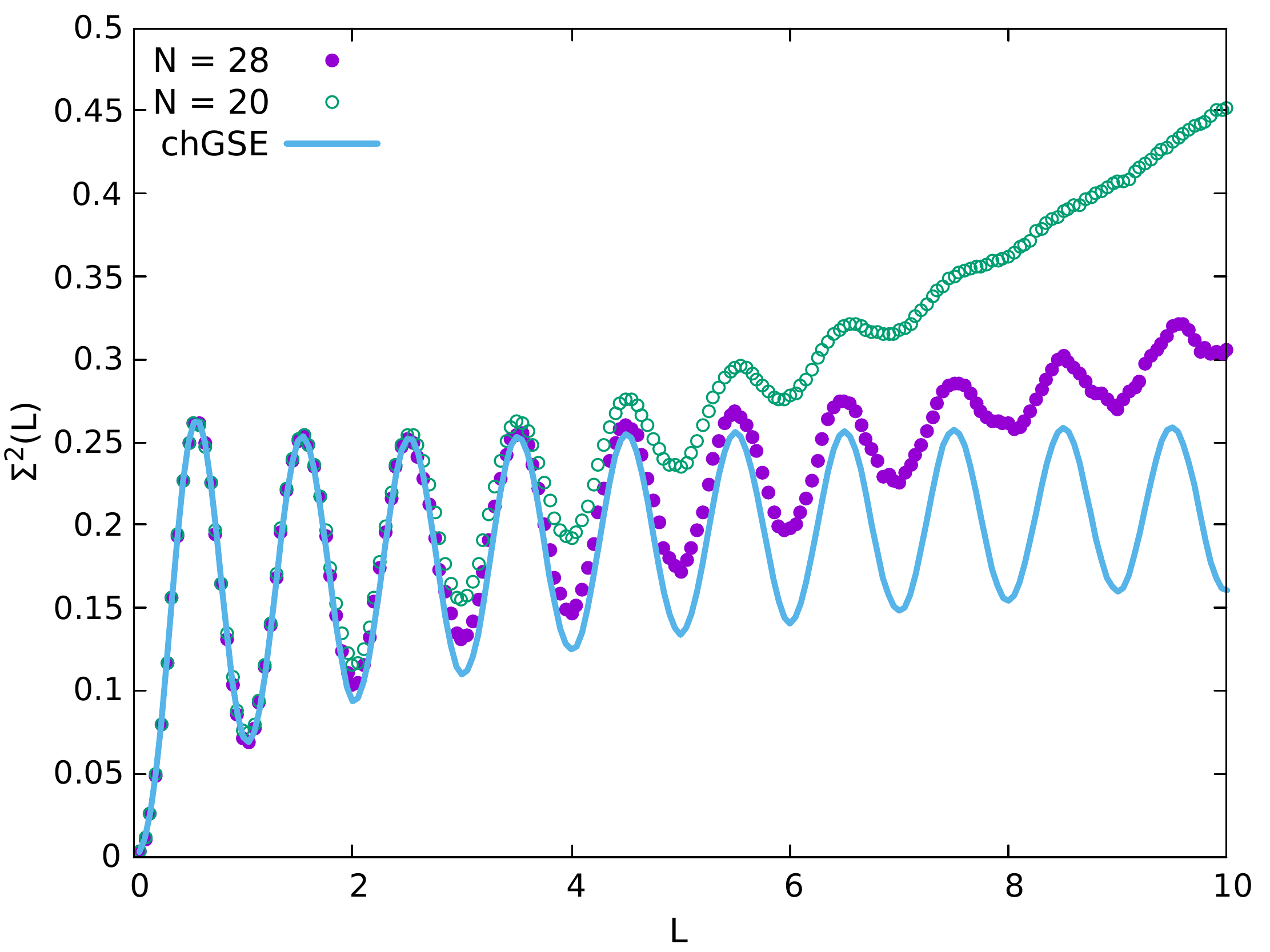}
     	\caption{Number variance $\Sigma^2(L)$ Eq.~(\ref{eq:nv}) for small $L$ and different $N$'s. Excellent agreement with random matrix theory is observed for sufficiently small $L$. This is an indication that SYK for odd $q$ is also quantum chaotic for sufficiently long times (small energies).
          The point in which deviations are observed, especially its scaling with $N$, provides useful information on the quantum dynamics prior to complete relaxation. Our numerical results are consistent with a point of departure that roughly scales with $N$.
     	}
     	\label{nvt}
     \end{figure}

 \section{Universality in level statistics and quantitative estimation of the Thouless energy}  
 We now proceed to the study of dynamical properties
 of the SYK model  by studying level statistics beyond the Thouless energy.
 In the non-supersymetric SYK model, it was found \cite{garcia2017,cotler2016} that
 in both the bulk of the spectrum and the region close to the ground state,
 spectral correlations for small eigenvalues separations are well
 described by random matrix theory. This is an indication that for
 sufficiently long times, of the order of the Heisenberg time,
 the SYK model is still quantum chaotic and has reached an ergodic
 state where the dynamics is universal since it depends only on the
 global symmetry of the system.  
 
 We focus again in the low energy region $E \approx 0$ where we expect
 more differences with respect to the SYK  model with $q$ even.  
 Since agreement with random matrix theory is expected for short-range
 spectral correlations, we investigate the number variance, a popular
 choice to characterize long range spectral correlations which will allow
 us a more systematic description of the corrections to the random matrix results.
 It is defined as the variance of the number of levels ${\tilde N}(\epsilon)$ in an energy
 interval of width $\epsilon$ (in units of the mean level spacing):
 \begin{equation}
 \Sigma^2(\epsilon) = \left\langle {\tilde N}^2(\epsilon)\right\rangle - \left\langle {\tilde N}(\epsilon)\right\rangle^2.
 \label{eq:nv}
 \end{equation}
 In order to proceed, we have to unfold the spectrum, namely, to rescale it
 so that the mean level spacing is energy independent. In addition
 we use the connected two-point function to study spectral correlations.
 The removal of the
 one point function contribution is crucial to relate results from
 level statistics to dynamical features of the system.
 
 We employ two unfolding methods: a polynomial of fifth degree fitting
 for the whole averaged spectrum and the splines method where consecutive
 groups of several 
 eigenvalues, at least of the order of the spectral window to be investigated, are fitted to a linear or quadratic polynomial. 
 Results from both methods are similar so we stick to the latter which
 provides a slightly more accurate fitting for the eigenvalues closest to the origin.  
 In terms of the unfolded spectrum  $\langle {\tilde N}(\epsilon)\rangle = \epsilon$,
 so $\langle {\tilde N}\rangle \equiv L$ where $L$ is the number of unfolded
 eigenvalues in the window of interest. The number variance can be expressed
 as the double integral of the connected two level correlation function whose analytical expression is known for all ensembles of random matrices
 \cite{nagao1995}. However, it is often simpler to generate these curves from numerical exact diagonalization
 of random matrices, which we will do 
 for comparison to the SYK model. 
 The number variance for  $L < 10$, starting from $E=0$,
is shown in Fig. \ref{nvt}
 for  different values of $N$, corresponding to different universality
 classes. In all cases we find excellent  agreement with the random matrix
 prediction for the first few eigenvalues.
 As was expected, because the SYK model is quantum chaotic, the agreement extends to more eigenvalues as $N$ increases.
 The point of departure from the  random matrix prediction, usually called Thouless energy, seems to scale as $N^\alpha$ with $\alpha \sim 1$ for all universality classes. This is consistent with  previous results for the microscopic spectral density.  However the value of $\alpha$, and therefore of the Thouless energy, for the non-supersymmetric SYK models ($q > 2$ even) seems to be much larger ($\alpha \approx 2$) \cite{garcia2016}.
 For $q$ even, there is
 an analytical argument that indeed $\alpha =2 $ \cite{altland2017}.  A reason for that behavior is that, for smaller $q$, the Hamiltonian is sparser and therefore it takes more time, so less energy, to explore the full phase space available. 
 
     \begin{figure}[t!]
     	\includegraphics[width=7.2cm]{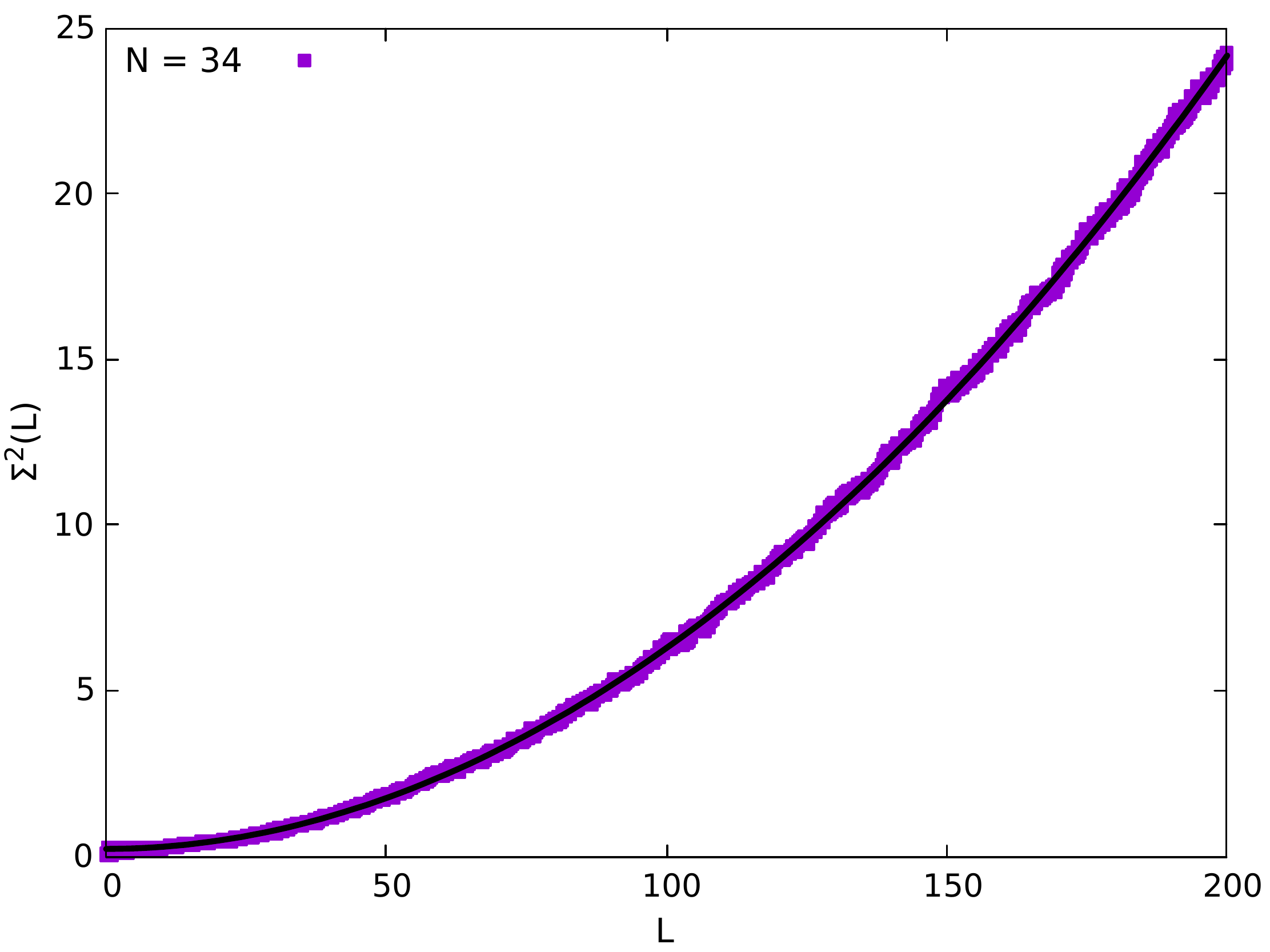}
     	\includegraphics[width=7.2cm]{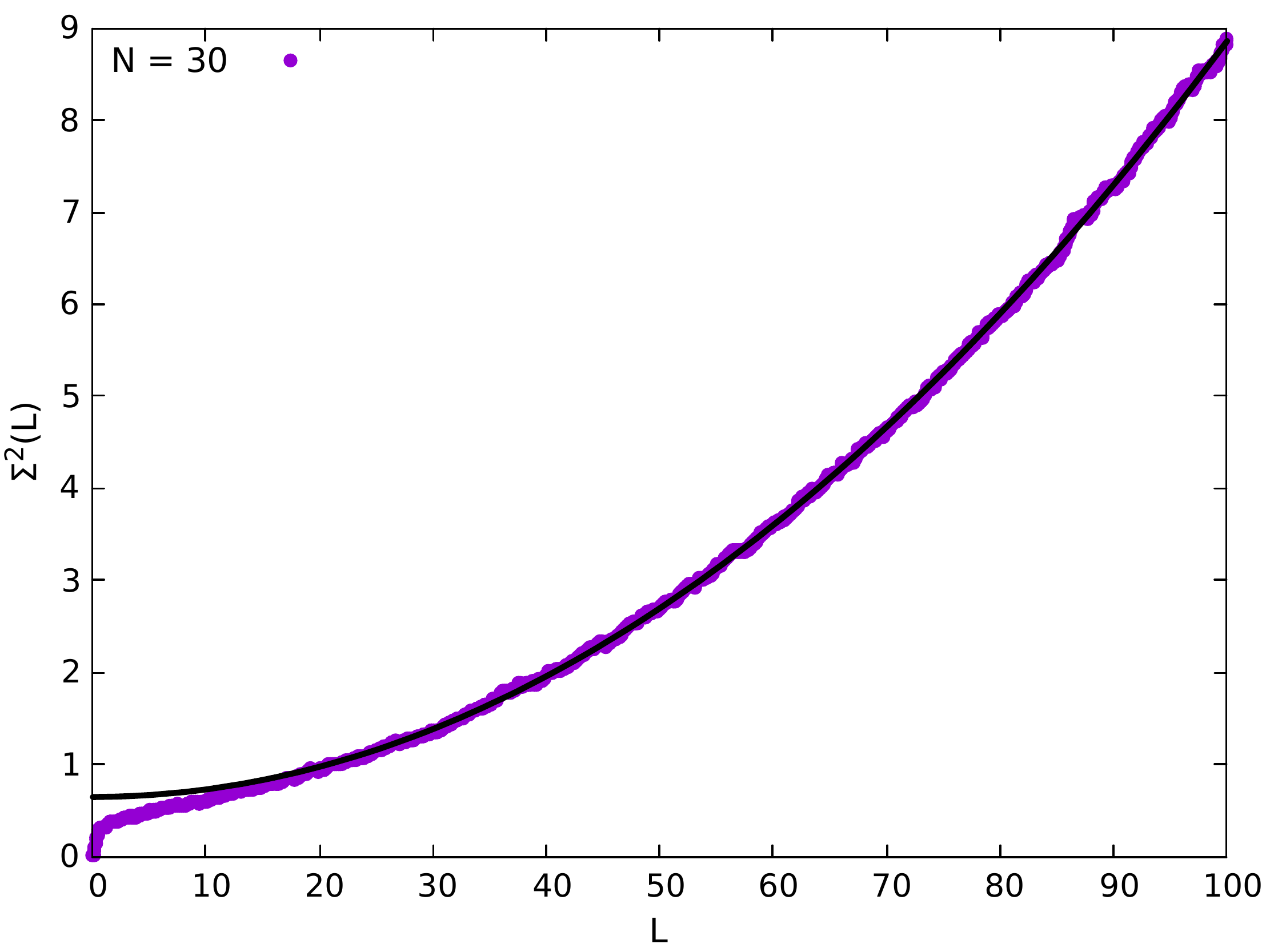}
     	\caption{Number variance $\Sigma^2(L)$ Eq.~(\ref{eq:nv}) for $N = 30$ and $N=34$. Deviations from the logarithmic growth predicted by random matrix theory are clearly observed.
     		We have fitted the growth (solid curves) to a function $\sim L^\alpha$. For $N$ sufficiently large $\alpha \approx 2$ ($\alpha \approx 1.99$ for $N=34$ and $\alpha \approx 1.98$ for $N=30$). For sufficiently small $N$, we find a bit slower growth but we believe that this is a finite size effect of no physical relevance. 
     	}
     	\label{nvthou}
     \end{figure}
     
 Beyond the Thouless energy, we observe a growth of the number
 variance much faster than the logarithmic growth predicted by random matrix theory.
 Our results, shown in Fig.~\ref{nvthou} for different values of $N$, are fully consistent with a quadratic growth $\sim L^2$ which is in agreement with the
 result obtained in Ref. \cite{altland2017}. 
 This quadratic growth can be understood as follows: in a nonlinear sigma model massive modes 
 contribute a constant $c_M$ to the connected two-level correlation function,
 \be
 \rho_{2,c}(\lambda_1, \lambda_2) = \rho_{2,c}^{\rm RMT} (\lambda_1,\lambda_2) + c_M
 -\delta \rho_2(\lambda_1,\lambda_2),
\label{twopan}
 \ee
 where the (short-range) correction $\delta \rho_2(\lambda_1,\lambda_2)$ is chosen such that
 the exact sum-rule is satisfied
 \be
 \int d\lambda_1\rho_{2,c}(\lambda_1,\lambda_2) = 0.
 \ee
 Note that $\rho_{2,c}^{\rm RMT}$ satisfies this sum-rule.
 The constant term in the two-point-function gives a quadratic contribution
 to the number variance. From Fig. \ref{nvthou} we find that the constant
 term in \eref{twopan} is well approximated by
 \be
 c_M \approx \frac \pi 4 \frac {1}{N^2}.
\label{cm}
 \ee
  The power law increase of the number variance is reminiscent of the result for weakly disordered metals $\sim L^{d/2}$ \cite{altshuler1986}, where $d$ is the spatial dimensionality of the
 system though it is not clear to us whether it is possible to push this analogy any further.

     \begin{figure}[t!]
     	\includegraphics[width=7.2cm]{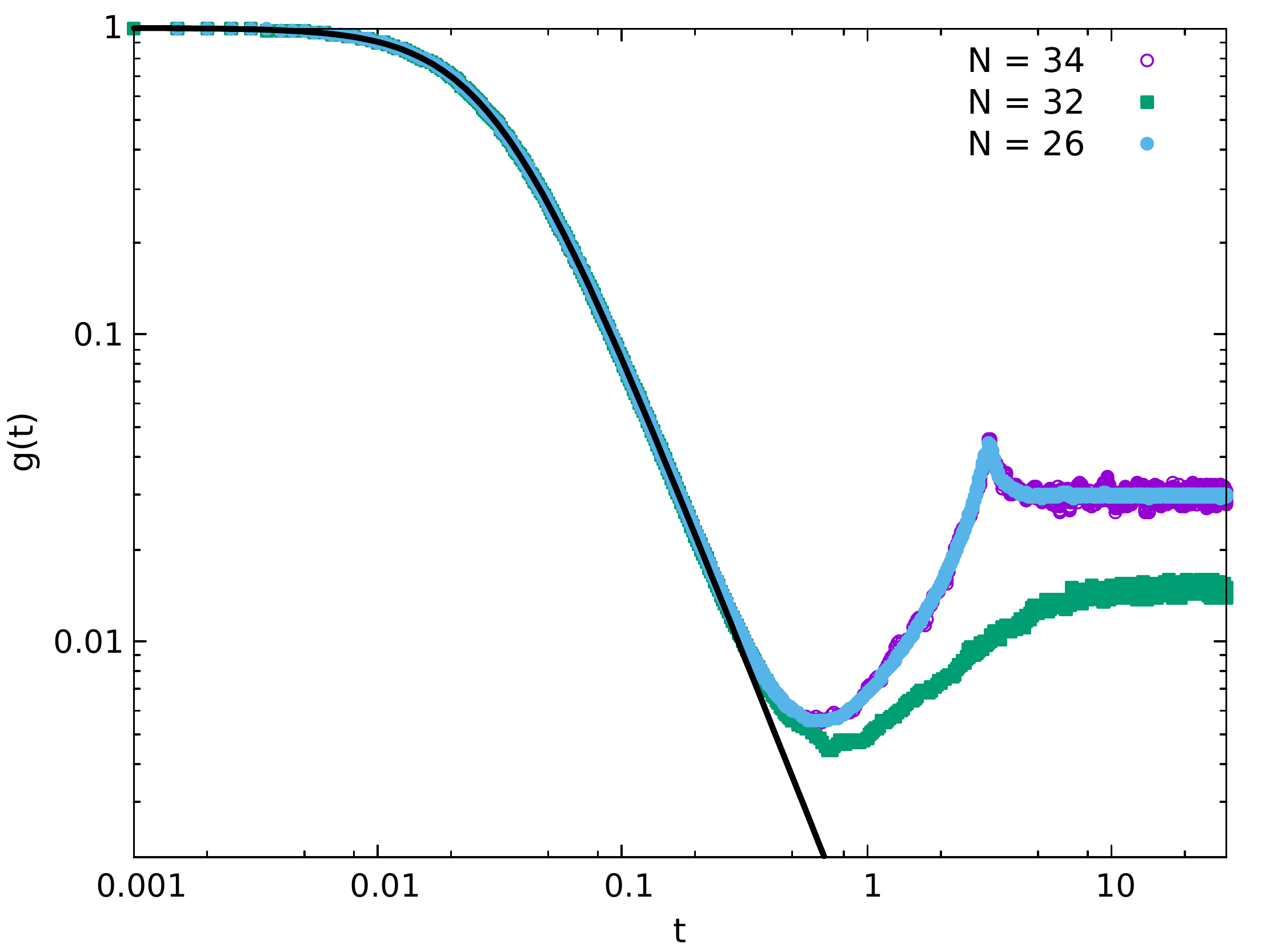}
     \includegraphics[width=7.2cm]{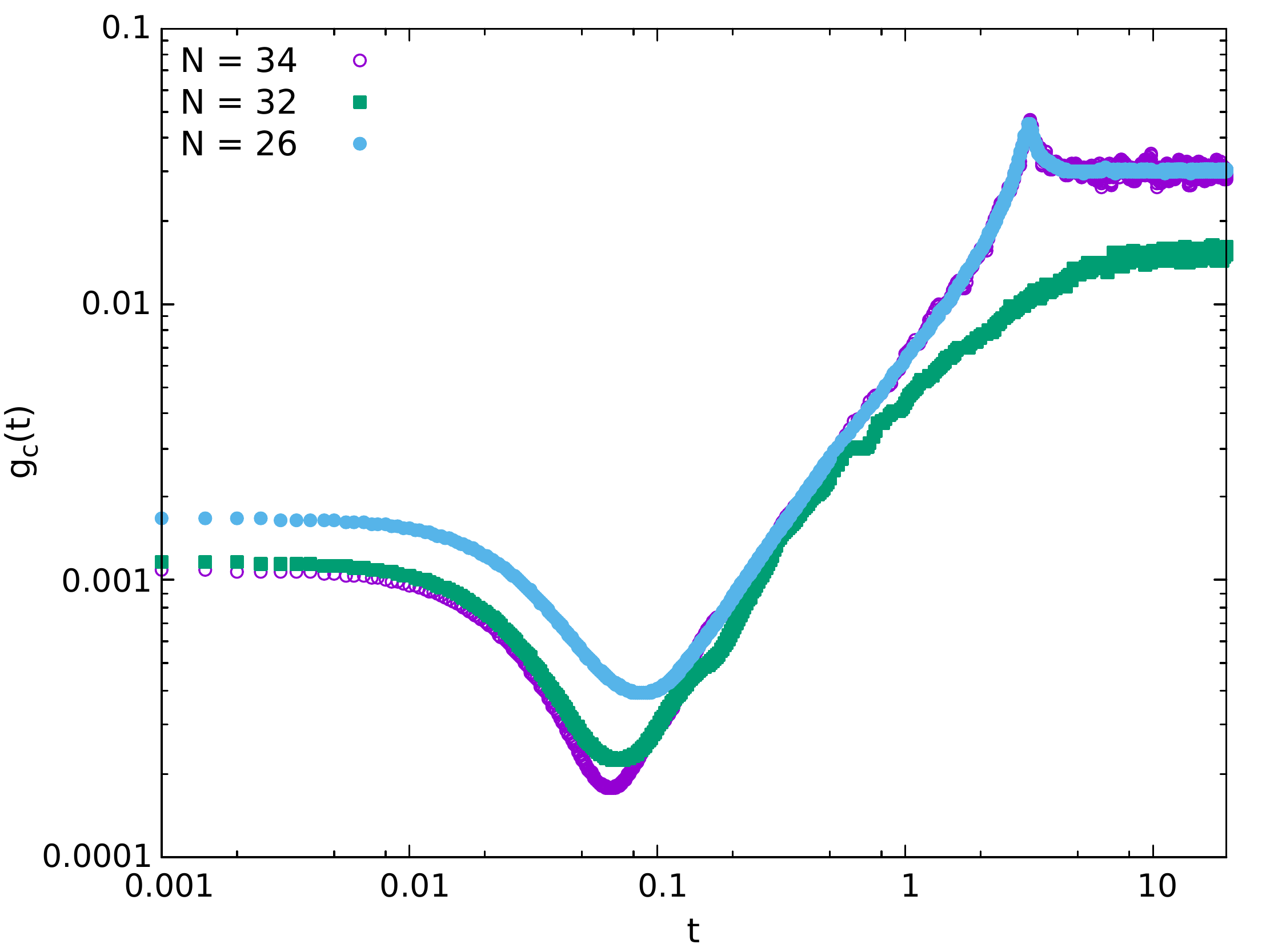}
     \includegraphics[width=7.2cm]{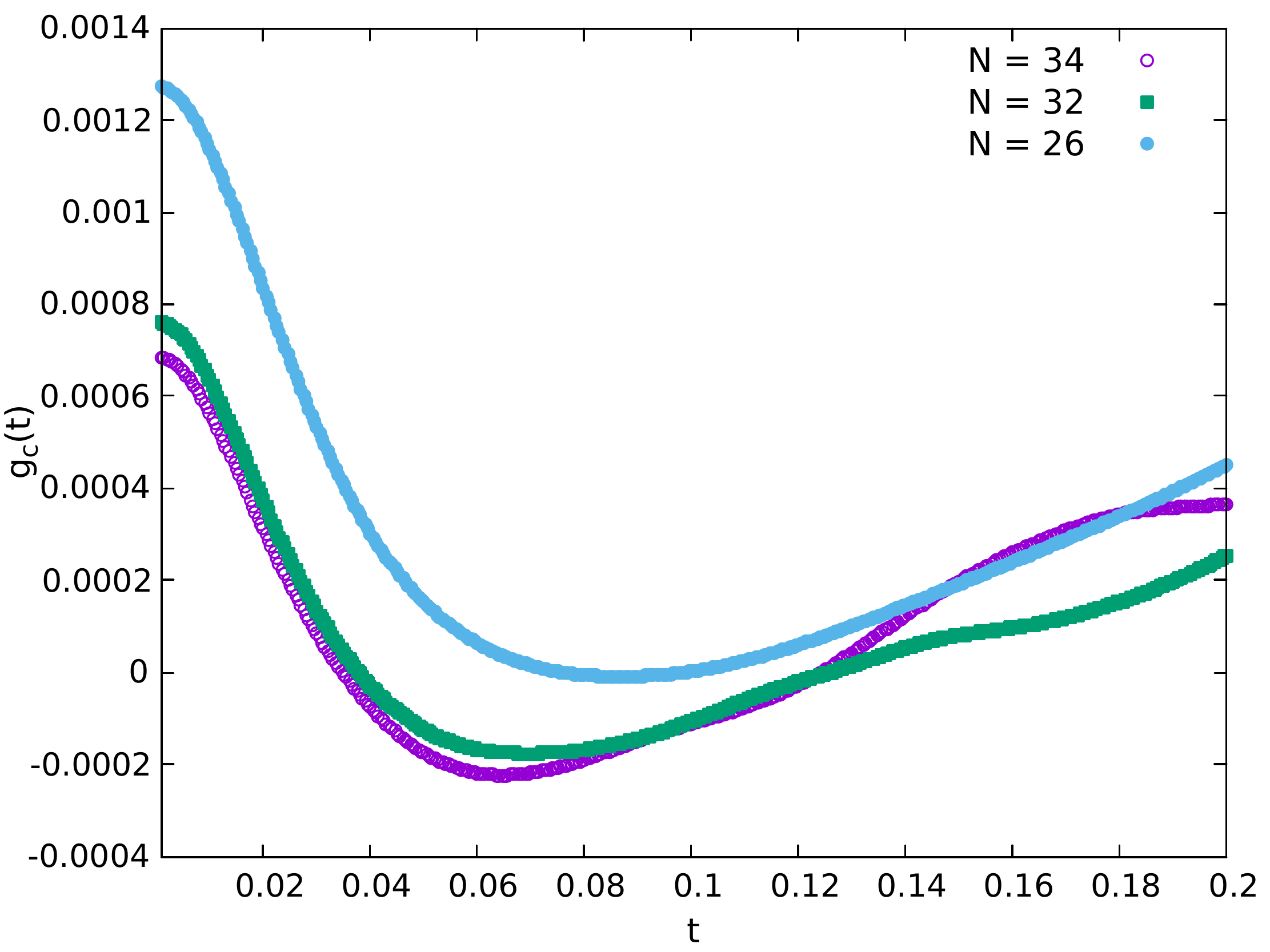}
      \includegraphics[width=7.2cm]{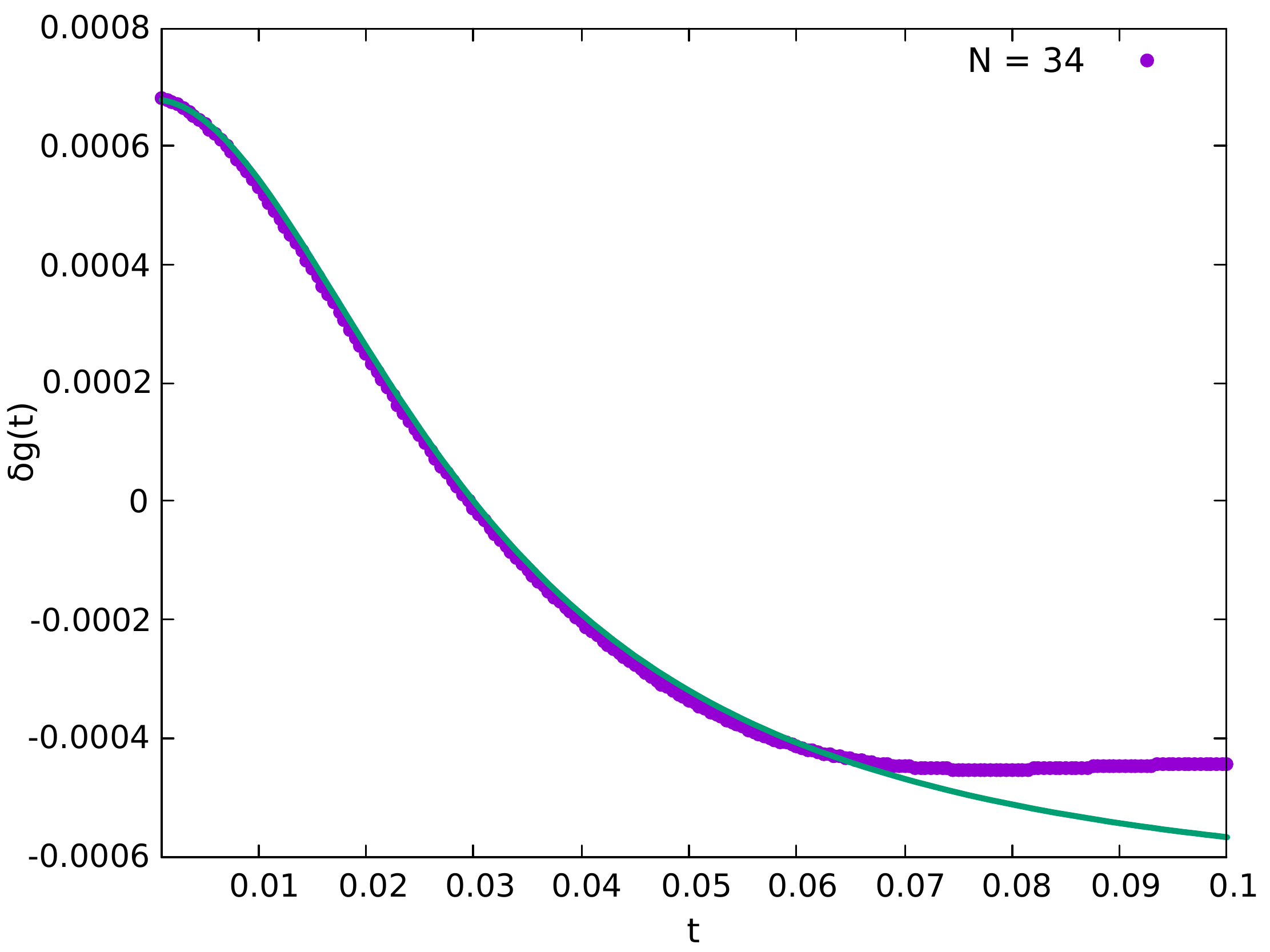}
     	
     \caption{Top-Left: Spectral form factor $g(t)$ Eq.~(\ref{eq:sff}), with $t$ in units of $J$ and $\hbar \equiv 1$,
       for $\beta = 0.03$ obtained from the unfolded spectrum of the
       supercharge Eq.~(\ref{hami}) and different values of $N$.
       The solid black curve represents the disconnected part of
       $g(t)$, $g_{d}(t)={\beta^2}/(\beta^2+t^2)$, the leading contribution
       to $g(t)$ for short times, that must be subtracted from $g(t)$ in order
       to extract the Thouless time $t_T$.
       Top-Right: Connected spectral form factor $g_c = g - g_d$.
       For sufficiently long times, we  observe a dip (correlation hole)
       at $t_T$ followed by a ramp and eventual saturation,
       all typical features of quantum chaotic system and also predicted
       by random matrix theory. The peak for $N =32$ and $N=34$
       is a known feature of quantum chaotic systems with symplectic symmetry.
       We have slightly shifted all curves by $0.0004$
       so that we can plot it in a log-scale.
       Bottom-Left: Zoom in of $g_c$ for short times.
       The position of the correlation hole seems to roughly scale with $N$.
Bottom-Right: the deviation from the random matrix theory result $\delta g(t)$  for $t < t_T$, before the dip is formed. Assuming fully chaotic eigenfunctions,  
 $\delta g(t)$ describes how the SYK model approaches equilibrium. We have found that the numerical result agrees well with the theoretical prediction Eq~(\ref{eq:intg}) (solid curve) which decays as $t^{-2}$,
     as $t_T$ is approached.  
     	}
     	\label{sfff}
     \end{figure}

 \subsection{Spectral form factor}
 
 In a time representation, the Thouless energy, and the subsequent quadratic growth of the number variance,
 are interpreted as the minimum time after which the dynamics is universal and how this universal regime
 is approached respectively. 
 In order to explore these time scales, we compute the spectral form factor \cite{alhassid1992,borgonovi2016,lea2017,Torres-Herrera2017} with the unfolded spectrum,
 \be\label{eq:sff}
 g(t) = \frac{\langle Z^*(t)Z(0)\rangle}{\langle Z(0)\rangle^2}
 \ee
 with $Z=\sum_ie^{i\lambda_it-\beta\lambda_it}$ with $\lambda_i$ the unfolded eigenvalues and $\beta$ the inverse temperature. We also remove the disconnected 
 part $g_d(t) ={\beta^2}/{(\beta^2+t^2)}$ related to the one-point function. 
 Although not always employed in the literature \cite{cotler2016}, it is necessary to work with unfolded eigenvalues and connected two-point functions for a correct determination of the Thouless energy. 
 
 
   
 We illustrate this point in Fig.~\ref{sfff} where we compute the full spectral form factor, for $\beta = 0.03$, together with the connected part $g_c = g - g_d$.  We observe the ramp and saturation expected from random matrix theory in both cases.
 We also observe a dip (correlation hole) for short times. Only for $g_c$ this is related to the Thouless time, a time scale the describes the limit of applicability of random matrix theory.
 The dip (correlation hole) occurs for $N =34$ at $t \sim 0.1$, about $1/10$
 of the Heisenberg time. 
 This is
 is roughly consistent with the estimation of the Thouless energy from the number variance 
 in Fig.~\ref{nvt} which was computed removing the double degeneracy (for $g(t)$ we kept the double degeneracy)  
and also using the unfolded spectrum and the connected part of the two-level correlation function.

The scaling of the $t_T$, see Fig.~\ref{sfff} (bottom left), seems to be linear with $N$, also consistent with the number variance estimation. For $t < t_T$, the spectral form factor deviates from the random matrix theory prediction.
Simple perturbation theory \cite{borgonovi2016,lea2017}, valid for $t \ll t_T$, shows that 
\be
\delta g (t) \equiv g_c(t) - g_c^{\rm RMT}(t) \sim \beta^2 - t^2.
\ee
Similarly, for $t \lesssim t_T$, the constant contribution $c_M$ in Eq.~\ref{twopan} controls $\delta g(t)$ so
  \be
  \delta g (t \lesssim t_T) \approx g(0) \frac{\beta^2}{\beta^2+t^2}.
  \ee
  with $g(0) \approx c_M$. More specifically, 
  for $N=30$ and $N=34$ the coefficient of the quadratic dependence
  of the number variance $c_M$ is $0.00080$ and $0.00062$, respectively, while $g(0)$,
  equal to $0.00075$ and $0.00068$.

In Fig.~\ref{sfff} (bottom right) we compare the numerical $\delta g (t)$ 
with a simple interpolating expression 
\be\label{eq:intg}
\delta g(t) = g(0) \frac{\beta^2 -t^2}{\beta^2+t^2},
\ee
valid for any $t < t_T$.
The agreement is excellent until times very close to $t_T$. 
 
As for the number variance, our results are reminiscent to those for
 non-interacting weakly disordered metals in more than two spatial dimensions where a power-law decay was observed \cite{altshuler1986,argaman1993}. However there are important differences. In our case, diffusion is in Fock space and it is unclear whether the techniques and ideas employed for a non-interacting problem can be translated to the interacting SYK model.  

 \section{Outlook, holographic interpretation and conclusions}
 We have studied spectral and thermodynamical properties of the
 supersymmetric SYK model ($q$ odd).
 We have found that the microscopic spectral density
 corresponding to the  $O(N)$ smallest eigenvalues is universal,
  and is well described by chiral or superconducting random matrix ensembles depending on the value of $N$.
 The average spectral density, computed analytically and numerically, 
 for $E \approx 0$ grows exponentially
 with $N$, though the chiral condensate, obtained from the spectral resolvent, which is normalized with respect to
 the total number of eigenvalues, vanishes in the
 thermodynamical limit.

 For slightly larger energies, the average spectral density grows
 exponentially. Spectral correlations in the
 region $E \approx 0$ also agree with the random matrix prediction for
 sufficiently short eigenvalue separations.
 For eigenvalue separation larger than the Thouless energy, which roughly scales with $N$,
 we observe deviations from random matrix theory which are characterized by a
 quadratic growth of the number variance. We have also found a simple analytical expression for the spectral form factor, valid for times shorter than the Thouless time,  that provides useful information about how the SYK model approaches ergodicity.

One of the main motivations to study the SYK model is its relevance in holography. The observation of quantum chaos for long
 times and the exponential growth of the low energy excitations together with a finite zero temperature entropy
 strongly suggest that this supersymmetric extension could also have a
 gravity dual. Moreover it also raises some interesting questions.
 For instance, 
 it would be interesting to understand in more detail the interpretation in the quantum gravity dual of the observed universal microscopic spectral density
or the existence of a larger entropy at zero temperature with respect to the non-supersymmetric case.

 \acknowledgments{Aurelio Berm\'{u}dez (A.M.G.),
   Bruno Loureiro (A.M.G.), Alex Kamenev (J.V.) and Lea Santos (J.V. and A.M.G) are thanked for illuminating discussions. Y.J. and J.V. acknowledge  partial support from
  U.S. DOE Grant
  No. DE-FAG-88FR40388.}

 \appendix

 \section{Spectral Density for Odd $q$}
\label{app:A}
 
In this Appendix we calculate the large $N$ limit of the spectral density
for odd $q$.

For negative $\eta$ the spectral density can be written as
\be
\rho(E) &=& c_N \sqrt{1-(E/E_0)^2 }\prod_{k=1}^\infty
  \left[1-4\frac{E^2}{E_0^2}\frac 1{2+|\eta|^{2k}+|\eta|^{-2k}} \right ]
  \left[1-4\frac{E^2}{E_0^2}\frac 1{2-|\eta|^{2k-1}-|\eta|^{-2k+1}} \right ]\nn \\
  &=& c_N \exp\left[\frac 12\sum_{k=-\infty}^\infty
      \log \left[1-4\frac{E^2}{E_0^2}\frac 1{2+|\eta|^{2k}+|\eta|^{-2k}} \right ]
\right . \nn\\  && \left.
      +\log\left[1-4\frac{E^2}{E_0^2}\frac 1{2-|\eta|^{2k-1}-|\eta|^{-2k+1}} \right ]
    \right].
\label{rhoqhodd}
\ee
With the normalization constant $c_N$ given by
\be
c_N = \frac {2^{N/2}}{\pi\sigma} (1+\eta)\sqrt{1-\eta}\prod_{k=1}^\infty \frac{1-\eta^{2k+2}}{1-\eta^{2k+1}}
\label{cn}
\ee    
the spectral density
  is normalized to $2^{N/2}$.
        After a Poisson resummation, the expression can be rewritten as
\be
        \rho(E) &=& c_N \exp \left [ \frac 12 \sum_{n=-\infty}^\infty \int dx e^{2\pi inx}
          \log\left [1 - \frac {E^2}{E_0^2}\frac 1{\cosh^2 x \log|\eta|}\right ]
          \right . \nn \\ &&\left .
          + \log\left [1 + \frac {E^2}{E_0^2}\frac 1{\sinh^2 (x-\frac 12)\log|\eta|}\right ]\right ].
          \ee
          Both integrals can be evaluated analytically
\be
 \frac 12 \int dx e^{2\pi inx}  \log\left [1 + \frac {E^2}{E_0^2}\frac 1{\sinh^2 (x-\frac 12) \log|\eta|}\right ] &=&\frac {(-1)^n}{2|n|} \left (1 - \exp \left [\frac{ 2\pi |n|}{\log|\eta|} |\arcsin (\frac{E}{E_0})| \right ] \right ) \label{oddn},\nn \\  
 \ee
 and
 \be
 \frac 12 \int dx e^{2\pi inx}  \log\left [1 - \frac {E^2}{E_0^2}\frac 1{\cosh^2 x \log|\eta|}\right ]& =&-\frac 1{2} \frac{1 - \cosh \left [\frac{ 2\pi |n|}{\log|\eta|} \arcsin (E/E_0) \right ]} {n\sinh(n\pi^2/\log|\eta|)}.
\ee
The case $n = 0$ has to be considered separately. After shifting $x$ by $1/2$ we obtain,
\be
\frac 12 \int dx   \log\left [1 + \frac {E^2}{E_0^2}\frac 1{\sinh^2 x \log|\eta|}\right ].
\ee
Using $x |\log\eta|$ as new integration variable, this can be written as
\be
&&-\frac 12 \frac 1{\log|\eta|} \int dx   \log\left [1 + \frac {E^2}{E_0^2}\frac 1{\sinh^2 x }\right ].\nn\\
&=&
+\frac 1{8\log|\eta|}  \left(\log(e^{-2i\phi})(-2\pi i \sign(\phi)+\log e^{2i\phi})
+\log (e^{2i\phi})(2\pi i \,\sign(\phi)  +\log e^{-2 i\phi}) \right )
\nn\\
&=& -\frac {\pi|\arcsin (E/E_0)|}{\log|\eta|}+
\frac{\arcsin^2(E/E_0)}{\log |\eta|}
.
\ee
where $\phi = \arcsin(E/E_0)$. Ignoring exponentially small contributions,
for $|E/E_0|< 1$  the terms with $n\ne 0$ in Eq. \eref{oddn} can be summed into
\be
\log ((1+e^{2\pi |\arcsin(E/E_0)|/\log |\eta|})/2) .
\ee

For large $N$ away from the edge of the spectrum the leading contribution
to the spectral density is dominated by the $n=0$ term resulting in the
asymptotic form for the spectral density
\be
\rho_{\rm asym} (E) = c_N  \cosh \left(\frac {\pi\arcsin(E/E_0)}{\log|\eta|}\right)\exp\left [ 2 
  \frac { \arcsin^2(E/E_0)}{\log|\eta|} \right ].
\label{rhoasyma}
\ee
The large $N$ limit of the normalization constant is determined by
\be
\int dE \rho_{\rm asym}(E) dE = 2^{N/2}.
\ee
For large $N$ the integral can be evaluated by a saddle point approximation. Using
that $\log| \eta| \sim  -2 q^2/N$ in this limit we find
\be
c_N = e^{N/2\log 2- N\pi^2/16 q^2},
\ee
which give exactly the leading order $1/q^2$ correction to the zero temperature entropy
\cite{fu2017}.

\section{The Q-Hermite Approximation for the Resolvent of the SYK Model}\label{app:res}

The resolvent of a Hamiltonian $H$ is  defined by
\be
G(z) = \frac 1{{\cal N}}{\rm Tr} \frac 1{z+H},
  \ee
  where ${\cal N} $ is the dimension of the Hilbert space.
If $|z/\lambda_k|>1$ for all eigenvalues $\lambda_k $ of $H$ it can be
expressed in terms of the moments of $H$. For a Hamiltonian for which
the odd moments vanish, such as the SYK model, we obtain
 \be
 G(z) = \sum_{p=0}^\infty \frac {M_{2p}}{z^{2p+1}}.
 \ee
This sum has a finite radius of convergence due to the fact
 that it has a cut along the support of the spectrum (for a Gaussian
 the radius of convergence is zero). However, it may be possible
 to analytically continue $G(z)$ inside the radius of convergence.
 This can be done using the resummation
 \be
 \frac 1{z^{2k+1}} = \sum_{j=0}^\infty (-1)^j {2k+j \choose j}
 \frac 1{(z/2+\sqrt{z^2/4-1})^{2(j+k)+1}}.
\label{an-con}
 \ee
   The sum is convergent on the imaginary axis. To prove this identity we start
from the binomial expansion 
\be
\frac 1{(e^x + e^{-x})^{2k+1}} =\sum_{j=0}^\infty {2k+j \choose j}e^{-x(2j+2k+1)}.
\ee
Defining $z$ as
\be
z = e^x +e^{-x},
\ee
$e^x$ is solved by
\be
e^x= \frac z2 \pm  \sqrt{z^2/4-1}
\ee
resulting in
\be
 \frac 1{z^{2k+1}} = \sum_{j=0}^\infty (-1)^j {2k+j \choose j}
 \frac 1{(z/2\pm\sqrt{z^2/4-1})^{2(j+k)+1}}.
\ee
For large $z$ only the positive sign in the r.h.s. reproduces the 
$1/z^{2k+1} $ term so that the correct analytical continuation is given 
by Eq. \eref{an-con}.

Because the spectrum of the SYK Hamiltonian is symmetric under $\lambda_k \to -\lambda_k$, the resolvent on the imaginary axis is purely imaginary and
\be
iG(is) = \frac 1{\cal N}\sum_{\lambda_k>0} \frac {2s}{\lambda_k^2+s^2}.
\ee
Using the analytical continuation  \eref{an-con} we obtain
   \be
   iG(is) &=& \sum_{p=0}^\infty (-1)^p \frac {M_{2p}}{M_2^p}\frac 1{s^{2p+1}}\nn\\
&=& \sum_{p=0}^\infty \frac{M_{2p}}{M_2^p}\sum_{j=0}^\infty (-1)^{p} {2p+j \choose j}
   \frac 1{(s/2+\sqrt{s^2/4+1})^{2(j+p)+1}}.
   \label{B9}
 \ee
 The second line of \eref{B9} also converges when $s$ is small.

 In the Q-Hermite case the  moments are given by
 \be
 \frac {M_{2p}}{M_2^p} =\frac 1{(1-\eta)^p} \sum_{k=-p}^p (-1)^k \eta^{k(k-1)/2}{2p \choose p+k}.
\label{momqhap}
 \ee
we will  show next that the result \eref{B9} can be simplified to
   \be
   iG(is) &=&  {\sqrt{1-\eta}} \sum_{k=0}^\infty 
   \frac { \eta^{k(k+1)/2}}
         {\left (\frac{s\sqrt{1-\eta}}{2}
           +\sqrt{ \frac {s^2(1-\eta)}{4}+1} \right )^{2k+1}}.
         \label{resvien}
         \ee
To prove this,         we first shift $s \to s\sqrt{1-\eta}$ which cancels against the prefactor
         of the Q-Hermite moments as can be seen from the first equation of
         Eq. \eref{B9}. Then we use the identity (see \cite{ismail1987})
\be
\frac 1{(\frac s2 + \sqrt{\frac{s^2}4+1})^{2k+1}}=\sum_{j=0}^\infty
  \frac {2k+1}{2k+2j+1} {2j+2k+1 \choose j} \frac 1{s^{2k+2j+1}}
  \ee
  to rewrite \eref{B9} as a double sum which can
  be rewritten as a sum over $p\equiv k+j$ fixed and a sum over $j$ which
  runs from 0 to $p$.  By splitting the sum over $k$ form $-p$ to $p$
in the expression for the Q-Hermite moments
  into a  
  sum from 0 to $p$ and a sum form $-p$ to $-1$ we see that the two
  expressions for the resolvent are the same.

   A scale can be re-introduced by the substitution
   \be
   iG(is) \to \frac{1}\sigma i G(is/\sigma).
     \ee

\bibliography{library2-6}

\end{document}